\documentclass[%
twocolumn,
 longbibliography,
 amsmath,amssymb,
 aps,
 pre,
]{revtex4-1}

\usepackage{mathtools}
\usepackage{lipsum}
\usepackage{graphicx}
\usepackage{dcolumn}
\usepackage{bm}
\usepackage{comment}
\usepackage{afterpage}
\usepackage{xcolor}
\definecolor{sapphire}{HTML}{0067A5}
\usepackage[colorlinks=true,citecolor=sapphire,linkcolor=sapphire,urlcolor=sapphire]{hyperref}

\begin{document}
\title{ {Rectification of Confined Soft Vesicles Containing Active Particles}} 
\author{M. C. Gandikota and A. Cacciuto}
\email{ac2822@columbia.edu}
\affiliation{Department of Chemistry, Columbia University\\ 3000 Broadway, New York, NY 10027\\ }

\begin{abstract}
\noindent One of the most promising features of active systems is that they can extract energy from their environment and convert it to mechanical work. Self propelled particles enable rectification when in contact with rigid boundaries. They can rectify their own  motion when confined in asymmetric channels and that of microgears. In this paper, we study the shape fluctuations of two dimensional flexible  vesicles containing active Brownian particles. We show how these fluctuations not only are capable of easily squeezing a vesicle through narrow openings, but are also responsible for its rectification when placed within asymmetric confining channels (ratchetaxis). We detail the conditions under which this process can be optimized, and sort out the complex interplay between elastic and active forces responsible for the directed motion of the vesicle  across these channels.
\end{abstract}

\maketitle
\section{Introduction}
It is well known that the second law of thermodynamics prohibits the extraction of work from any cyclic process performed on a system in thermal equilibrium with a single thermal bath~\cite{leighton1965}. This exploit can instead be achieved in systems that are subject to active fluctuations. Seminal experiments on  micro-gears placed in a suspension of bacteria have demonstrated how sustained and rectified rotational motion of a gear can be achieved~\cite{angelani2009self,di_leonardo_bacterial_2010,sokolov_swimming_2010}.
Further work demonstrated how motion rectification can occur when \textit{E. coli} is confined in a chamber divided by funnel shaped walls~\cite{galajda_wall_2007,wan_rectification_2008,ratchet_review} and when active components such as active colloids or chains are subject to  asymmetric potentials~\cite{angelani2011active,yariv2014ratcheting,chen2015ratchet} or when confined in  asymmetric periodic channels~\cite{ghosh2013self,ao2014active,lu2017ratchet}.  

Motion rectification  breaks both space and time symmetries  and is thus not permitted in equilibrium systems except as transients. In contrast, rectification can be achieved in thermal systems using time-dependent spatially asymmetric potentials~\cite{reimann2002brownian,hanggi2009artificial}.
Time-dependent potentials are, however, unnecessary for active particles to achieve rectification of motion given that active systems are inherently out of equilibrium at the single particle level~\cite{ramaswamy_mechanics_2010,marchetti_hydrodynamics_2013,zottl_emergent_2016}.

While previous studies on active rectification have mostly focused on active components directly interacting with hard passive objects, be that a gear or a micro-channel, here we show how rectification can also develop as a result of the interplay between active and elastic forces. We report the rectification of  soft vesicles enclosing active Brownian particles when confined in an asymmetric periodic channel. 

Experimentally, the vesicle can be realized by enclosing active particles inside lipid membranes, and it has been shown that large non-equilibrium shape deformations capable of reshaping the vesicle can develop under certain conditions~\cite{vutukuri2020active}. Experiments on macroscopic soft circular scaffolds that enclose small centimeter size actuators  have also been shown to be capable of squeezing through tight spaces~\cite{boudet2021}. Overall, the shape and motility aspects of these vesicles have been studied in detail by employing simple models such as a closed flexible boundary enclosing spherical or rod shaped active particles~\cite{paoluzzi2016shape,wang2019shape,quillen2020boids}. However, to the best of our knowledge, rectification aspects of vesicles undergoing active fluctuations have not been paid any attention. Yet this is an important problem especially given the experimental demonstration of directional motion of fibroblasts and epithelial cancerous cells when confined within asymmetric periodic channels~\cite{le2020ratchetaxis}. While cells are complicated entities involving sub-structures such as the cytoskeleton, here, we show that rectification can be achieved by far simpler deformable entities such as a two dimensional soft vesicle that encloses spherical active particles.

\section{Model}
We model the two dimensional vesicle as a flexible closed boundary consisting of $N$ self-avoiding passive spheres of diameter $\sigma$ connected with stiff harmonic springs. The active fluctuations of the vesicle are due to $N_A$ active Brownian particles (ABP) of spherical shape and diameter $\sigma$ enclosed within its perimeter. 
These particles are activated using a self-propelling force of constant magnitude $v_p$ and follow these equations of motion:
\begin{equation}\label{langevin}
	\begin{split}
		\frac{d\pmb{r}_i(t)}{dt}  &=  \frac{1}{\gamma} \pmb{f}(\{r_{ij}\}) +   v_p \,  \pmb{\hat{n}}_i(t)\,\delta_{t_i,1}  + \sqrt{2D}\,\pmb{\xi}(t),\\
		\frac{d \pmb{{\hat{n}}}_i(t) }{dt}&=\sqrt{2D_r}\, \pmb{\xi}_r(t) \times \pmb{\hat{n}}_i(t),
		\end{split}
		\end{equation}
where  $i$ is the particle index, $\pmb{\hat{n}}$ is the axis of propulsion, and $t_i$ is a binary index, which is set to zero (one) for passive (active) particles. The Kronecker delta function $\delta_{t_i,1}$ activates the self-propulsion term for the active particles. 
 The translational diffusion coefficient $D$ is related to the temperature $T$ and the translational friction $\gamma$ via the Stokes-Einstein relation $D=k_{\rm B}T\gamma^{-1}$. Likewise, the rotational diffusion coefficient, $D_r=k_{\rm B}T\gamma_r^{-1}$, with $D_r = 3D\sigma^{-2}$. The solvent-induced Gaussian white-noise terms for both the translational $\pmb{\xi}$ and rotational $\pmb{\xi}_r$ motions are characterized by $\langle \pmb{\xi}(t)\rangle = 0$ and $\langle \xi_m(t) \xi_n(t^\prime)\rangle = \delta_{mn}\delta(t-t^\prime)$. $\pmb{f}(\{r_{ij}\})$ indicates the excluded volume forces between particles and the stretching forces between neighboring monomers of the vesicle.
 Excluded volume forces between any two particles are enforced via a Weeks-Chandler-Andersen (WCA) potential
 $U_w=4\varepsilon\left[ \left( \frac{\sigma}{r_{ij}}\right)^{12} - \left(\frac{\sigma}{r_{ij}}\right)^{6} +\frac{1}{4}\right ].$
The bonds between neighboring monomers obey the harmonic potential  $U_s=k_s(r_{i,i+1}-\sigma)^2$ where $r_{i,i+1}$ is the distance between consecutive monomers along the chain. The spring constant $k_s$ is set to $2500 k_{\rm{B}}T/\sigma^2$.

\begin{figure}[t]
\centering
\includegraphics[width=0.4\textwidth]{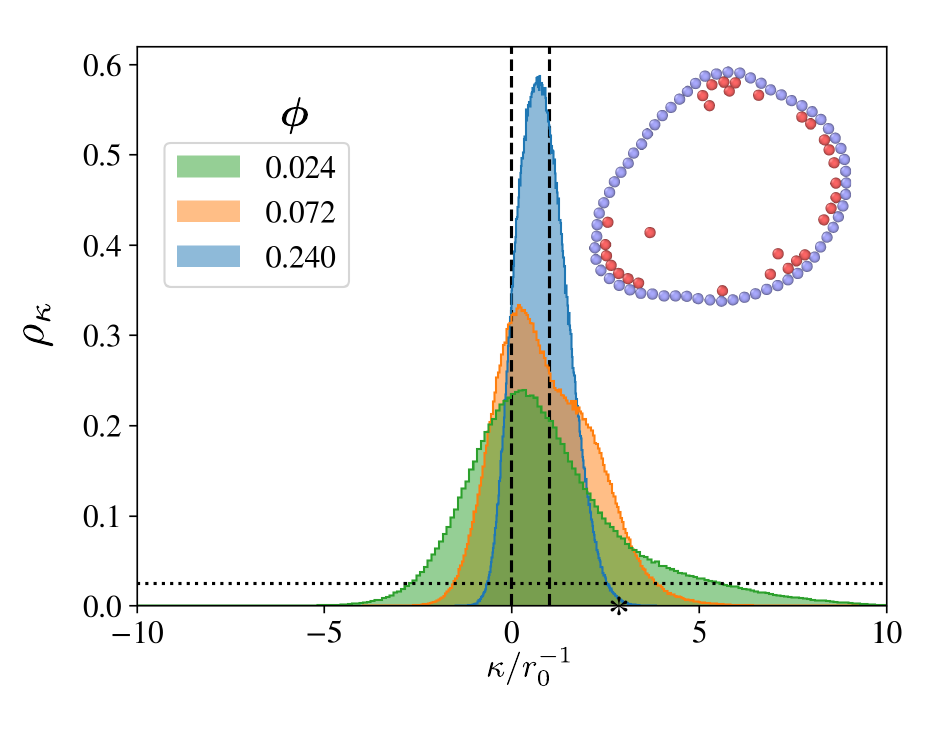}
 \caption{Probability distribution of local curvatures $\kappa$ for different numbers of ABPs. The vertical dashed lines are guides to the eye indicating $\kappa=0$ (flat region) and $\kappa=r^{-1}_0$ (the curvature of the exact circle described by the vesicle). The $*$ sign on the x-axis is the pore curvature in the circular cavities. The inset is a typical snapshot from our simulations with $\phi=0.072$. The flat distribution of curvatures of the passive ring ($N_A=0$) is shown as a black dotted line.} 
 \label{local curvature}
\end{figure} 

 We performed our simulations using the numerical package LAMMPS~\cite{plimpton_fast_1995} and set $\sigma$, $\tau=\sigma^2D^{-1}$ and $k_{\rm B}T$ as the units of length, time  and energy respectively. All simulations were run with a time step of $\Delta t=10^{-5}\tau$. We collected statistics for up to $10^{10}$ time steps. The strength of the active forces is reported in terms of the P\'eclet number defined as $Pe=v_p\sigma/D$. The packing fraction of the ABPs $\phi=N_A/(\pi r_0^2)$ where $r_0$ is the radius of the exact circular shape of the soft vesicle. We used a soft vesicle composed of $N=64$ colloidal particles unless otherwise stated.
 
\section{Shape fluctuations and escape from a circular cavity}

 We first study the effect of the active particles on the shape of a soft vesicle in open boundary conditions.

For an inextensible soft vesicle, the shape statistics of the vesicle is fully determined by the distribution of the local curvatures. We fitted osculating circles to every three consecutive monomers and determined the local curvatures by taking the inverse of their radii. The probability distribution of the local curvatures $\rho_k$ for varying number of active particles, $\phi$, at $Pe=100$ is shown in Fig.~\ref{local curvature}. For large $\phi$, the particles distribute homogeneously along the vesicle perimeter and push on it. The imposed tension on the vesicle suppresses the shape fluctuations~\cite{vutukuri2020active,iyer2022} resulting in a rather sharp curvature distribution centered around $\kappa=r_0^{-1}=0.098 \sigma^{-1}$. For small $\phi$, particles create local kinks on the vesicle and stretch flat the region of the vesicle connecting any two kinks. This results in a broad distribution of curvatures that peaks at $k/r_0^{-1}\simeq 0$ and extends all the way to $\kappa/r_0^{-1}=10$. For intermediate values of $\phi$, the distribution presents a peak near $\kappa/r_0^{-1}=0$ and develops a shoulder (rather than a distinct second peak observed for run-and-tumble particles~\cite{paoluzzi2016shape}) for larger values of $\kappa$, suggesting the presence of moderately curved regions generated by particle clustering on the surface that are then connected by flat regions (see a typical snapshot in Fig.~\ref{local curvature}.) Curvature-mediated particle clustering is a well known phenomenon for ABP near a curved surface~\cite{fily_dynamics_2014}, and plays an important role in understanding the active shape  of these flexible boundaries.

\begin{figure}[t]
\centering
\includegraphics[width=0.4\textwidth]{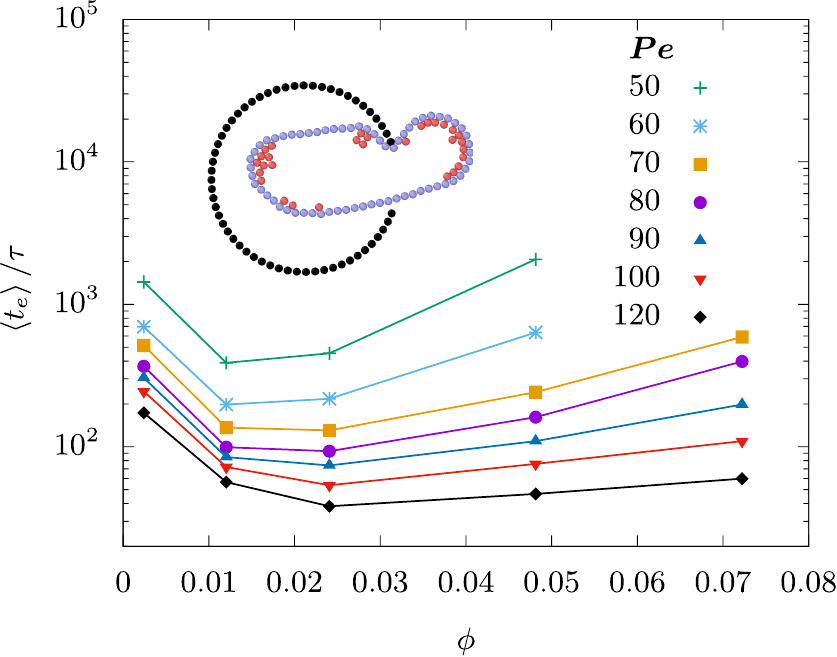}
 \caption{Average escape time of a soft vesicle from a circular cavity as a function of number of ABPs, $N_A$ for different values of $Pe$. The inset is a typical snapshot from our simulations for $\phi=0.072$ at $Pe=100$.}
 \label{escape time}
\end{figure} 

The presence of these large curvature fluctuations is of crucial importance when considering the ability of vesicles  to squeeze through narrow openings, as they 
 can be exploited to drive the vesicle out of a confinement cavity with a narrow opening. We study this phenomenon by considering a circular hard wall constructed by freezing in space a number of WCA particles of diameter $\sigma$ as shown in the schematic of Fig.~\ref{escape time}. The wall has an opening of size $7.9\sigma$ which allows the passage of the vesicle only if the local curvature of the vesicle at this pore is greater than $1/7.9\sigma=0.25\sigma^{-1}$. This curvature is marked with a ``$\ast$" on the $\kappa/r_0^{-1}$ axis of Fig.~\ref{local curvature}, and indicates how vesicles containing small numbers of active particles will be more likely to generate  large enough shape fluctuations to drive the vesicle out of the cavity.

 \begin{figure}[t]
\centering
\includegraphics[width=0.4\textwidth]{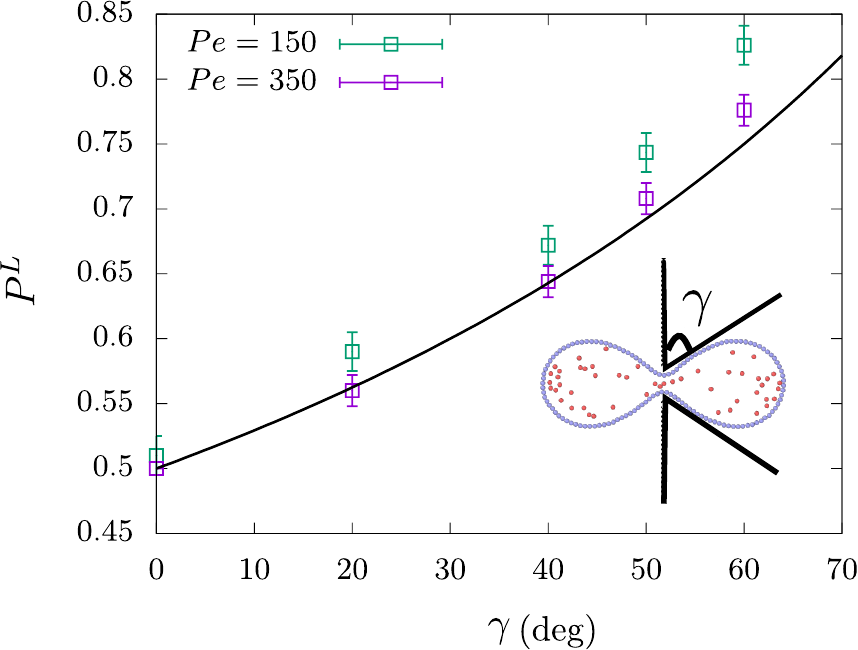}
 \caption{Establishing a preferential direction of motility using asymmetric walls. The vesicle escapes to the left for all finite $\gamma$ in the wall geometry specified in the schematic. The solid curve is Eq. (\ref{pressure diff}). Here, $\phi=0.03$ and $Pe=150$.}
 \label{premot fig}
\end{figure} 

The average escape time $\left<t_e\right>$  is calculated over a hundred runs for a range of $Pe$ and $\phi$ and is shown in Fig~\ref{escape time}.  Interestingly,
the escape time has a distinctly non-monotonic dependence on $N_A$. This can be understood by considering that while  small number of active particles lead to large curvature fluctuations of the vesicle, larger values of $\phi$ result in overall larger driving forces and cooperative pushing on the vesicle (see snapshot in Fig.~\ref{escape time}). However, if the vesicle is filled with too many active particles, curvature fluctuations become rare, and the vesicle escape is hindered. We find that $\phi\in[0.012,0.024]$ provides the best compromise between these two tendencies for the range of Pecl\'et numbers considered in this study.

\section{Rectification in asymmetric periodic channels}
We have now seen the interplay of the high-curvature segments of the vesicle and the directed motility of active particles which squeeze the vesicle through a pore. However, a vesicle containing active particles in free space does not have a preferential direction of motion. The mean square displacement (MSD) of the center of mass of this system can be calculated to be (see Appendix A)~\cite{paoluzzi2016shape,hagen2009non},
 \begin{equation}\label{msd1}
\left<\bm{r}^2(t)\right>=\frac{4D\;t}{N+N_A}+\frac{2\;Pe^2\sigma^2N_A}{9\;(N+N_A)^2}\left(e^{-D_r t}+D_rt-1\right).
\end{equation}
 The vesicle is diffusive in the limit of $t \gg \tau$ with the long time diffusion constant attaining its maximum for $N=N_A$ (see Appendix \ref{msd_appendix}). The MSD of a vesicle containing active particles can be considered to be the MSD of a single active particle with re-scaled diffusion constant and P\'eclet number while keeping the rotational diffusion constant unchanged. 

 For the vesicle to achieve directional motion, we need to break the spatial symmetry in the system~\cite{reimann2002brownian}. For instance, we can constrict the vesicle between two vertical walls as shown in the schematic of Fig.~\ref{premot fig}, where the active particles are equally distributed between the left and right regions. With this setup, the vesicle has equal probability of escaping towards the left or the right side of the space, i.e. $P^L=P^R=0.5$. The probabilities are computed by calculating the number of times the vesicle translocates to the left (or right) divided by the total number of translocation events. A simple way of breaking the spatial symmetry is to add a second set of walls tilted of an angle $\gamma$ with respect to the vertical axis as depicted in Fig.~\ref{premot fig}.  Crucially, as shown in Fig.~\ref{premot fig}, for any $\gamma>0$, $P^L>0.5$ and it monotonically increases with $\gamma$. For large $Pe$ and $N_A$, the vesicle conforms to the shape of the walls with most of the active particles located along its perimeter with their axis of propulsion parallel to the surface normal. In this limit one can estimate the relative forces pushing the vesicle on either side. While all particles on the left hand side contribute to push the vesicle their way by applying an homogeneous internal pressure to the left lobe, particles along the slanted walls of the right hand side generate no contribution to the overall pressure on the right lobe. In this limit, $P^L(\gamma)$ can be approximated to be (see Appendix B),
\begin{equation}\label{pressure diff}
P^L(\gamma)=\frac{1}{2(1-\gamma/\pi)},
\end{equation}
and is shown as the solid curve in Fig. \ref{premot fig}.

\begin{figure}[t]
\centering
\includegraphics[width=0.4\textwidth]{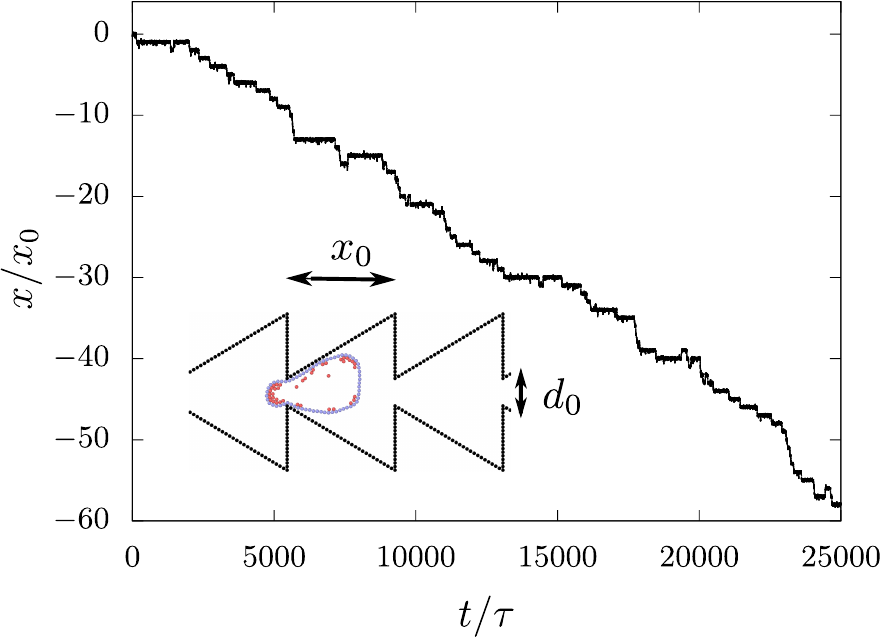}
 \caption{The position of the vesicle's center of mass denoted by $x$ as a function of time $t$. The vesicle displays leftward rectification when confined within a periodic asymmetric channel  The inset is a snapshot from simulations. Here, $\phi=0.072$ and $Pe=150$.}
 \label{rat fig}
\end{figure} 

We next consider a channel constructed with periodic cavities incorporating such a sloped geometry as shown in Fig.~\ref{rat fig}. The cavity is characterized by a width $x_0=30\sigma$, a pore size $d_0=7.6\sigma$ and  $\gamma=\text{atan}(0.6)$.

\begin{figure*}[t]
\centering
\includegraphics[width=0.8\textwidth]{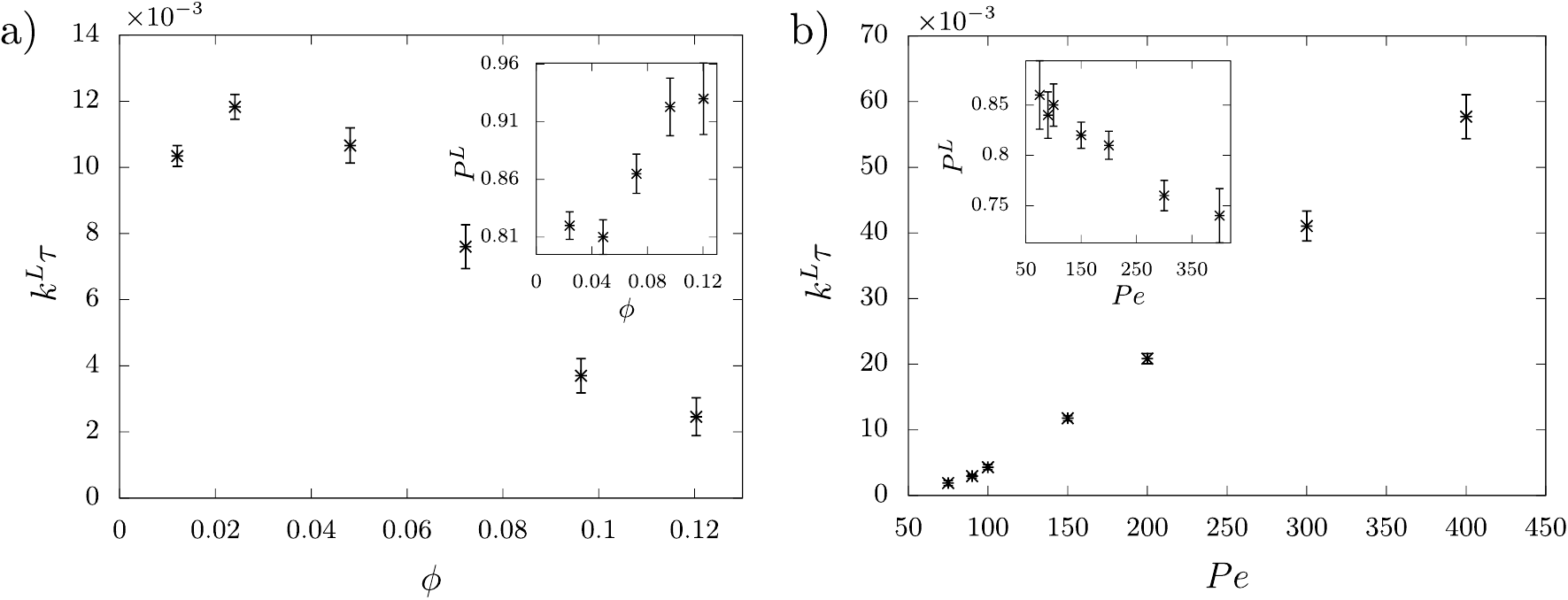}
 \caption{
a) The leftward escape rate of vesicle $k^L$ is non-monotonic with respect to $N_A$. Inset: The probability of leftward transitions consistently increases with the number of ABPs. Here, $Pe=150$. b) The leftward escape rate monotonically increases with $Pe$ while the probability of leftward transitions monotonically decreases (inset). Here, $\phi=0.024$.}
 \label{rate fig}
\end{figure*} 

The preferential leftward motion of the center of mass of the vesicle, $x$, is readily observed and highlighted in  Fig.~\ref{rat fig}. The rectification of motion as measured by $P^L$ consistently increases with $\phi$ as seen in the inset of Fig.~\ref{rate fig}(a). However, the escape rate from the narrow (left) side of the cavity in the periodic system, $k^L$, is non monotonic as seen in Fig.~\ref{rate fig}(a). The escape rate has its maximum at $\phi\approx0.024$ and decreases with further increase in $\phi$ consistent with the minimum of the average escape time in Fig.~\ref{escape time}. The escape rate monotonically increases with $Pe$ while the rectification of motion monotonically decreases as can be seen in Fig.~\ref{rate fig}(b) and its inset. Clearly, for $Pe=0$, that is when the vesicle is passive, we do not expect any rectification and expect $P^L=0.5$. However, measuring $P^L$ for $Pe<50$ is a challenge as each translocation event becomes extremely slow making it very hard to collect sufficient statistics.

\begin{figure}[h]
\centering
\includegraphics[width=0.4\textwidth]{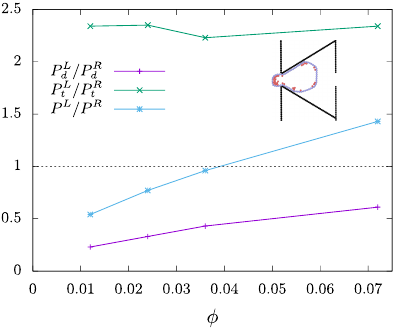}
 \caption{Relative docking and translocation probabilities for a vesicle confined inside an isolated triangular cavity for different values of $\phi$. The inset is a snapshot of a vesicle in this cavity. The lines connecting the data points are a guide to the eye.}
 \label{single_cavity}
\end{figure}

We have already identified one mechanism that explains why the left side of the cavity is the more likely  escape route for the vesicle.
There is a secondary driving force that becomes important for intermediate to large number of particles and results from the collective tendency of active particles to cluster around highly curved regions. To understand how this comes about, we considered a non-periodic, isolated triangular cavity as depicted in Fig.~\ref{single_cavity}. Within this cavity, the vesicle faces no asymmetry immediately upon exiting the cavity. If we factor the escape probabilities into the probability of docking, $P_d$, and that of translocating, $P_t$, then the ratio between $P_L$ and $P_R$ can be written as
\begin{equation}
\frac{P^L}{P^R}=\left(\frac{P^L_d}{P^R_d}\right)\left(\frac{P^L_t}{P^R_t}\right).
\end{equation}
Our simulations on this system indicate that the docking probability ratio $P^L_d / P^R_d$ increases with $N_A$ whereas the transition probability ratio $P^L_t / P^R_t$ remains roughly independent of $N_A$. See Fig.~\ref{single_cavity}. Crucially, in this cage, we observe a cross-over behavior where for small values of $\phi$, the vesicle rectifies its motion to the right, whereas for larger values of $\phi$, leftward rectification is achieved. This result suggests that for small number of particles, when the vesicle can undergo large curvature fluctuations, the slanted side of the cage effectively acts  as a  barrier against the vesicle passage by reducing the amount of space available to it. However, when $\phi$ becomes sufficiently large, and large shape fluctuations become rare, the slanted side of the cavity can promote the formation of high curvature regions (as the vesicle conforms to the shape of the walls) which in turn drive the accumulation of active particles that can cooperatively push the vesicle  through the left side pore. Clearly for large $\phi$ the docking probability decreases for both sides of the cavity, however, the decrease is more pronounced on the right side where such cooperative behavior is harder to achieve. Since,  as was already established above (see Eq.~\ref{pressure diff}), the probability of translocation is more likely on the slanted side of the cavity. For about $\phi>0.036$ the overall rectification direction switches from right to left.

\section{Relevance of vesicle size for rectification}
Lastly, we consider the behavior of a vesicle placed in a periodic asymmetric channel with a single active particle in its interior. 
Individual active particles are known to rectify their motion in asymmetric channels when the persistence length of the particles is larger than the cavity size (Knudsen regime)~\cite{ghosh2013self}. Operating in the Knudsen regime for the isolated particle ($Pe=100$), we find that while a particle in the absence of the vesicle readily rectifies its motion, when we place it inside the vesicle the response of the vesicle is crucially dependent on the vesicle 
size. Vesicles with a diameter $2r_0$ smaller than the pore size $d_0$ do not rectify their motion. However, for diameters $2r_0>d_0$,  vesicles rectify to the left (see table \ref{tab pl}). 

One could expect that placing an active particle within a small vesicle ($2r_0<d_0$) could move it away from the Knudsen regime because the effective P\'eclet number of the particle-vesicle system decreases with vesicle size according to $Pe\rightarrow Pe/(1+N)$ (see Eq. \ref{msd1}). However, the rectification of larger vesicles under the same conditions is counter intuitive. In fact, if anything, one should expect the  particle-vesicle system to further reduce its Pe\'clet number upon increasing the size of the vesicle ($N$). The emergence of $d_0$ as a relevant lengthscale for the problem  is an important result because it shows that the rectification of the vesicle is not a simple consequence of the rectification of the individual particles, but emerges from a complex interplay between active and elastic forces. 

\begin{table}[h]
\small
  \caption{\ Probability that a vesicle containing a single active particle with $Pe=100$ escapes from the left side of an asymmetric channel for different ratios of vesicle size, $2r_0$ to pore size, $d_0$}
  \label{tab pl}
  \begin{tabular*}{0.48\textwidth}{@{\extracolsep{\fill}}l|lllll}
$2r_0/d_0$ &0.27& 0.55 & 1.09& 1.64& 2.18\\
\hline
$P^L$      &0.50 & 0.50 & 0.50& 0.53& 0.62 \\    
  \end{tabular*}
\end{table}

This can be qualitatively understood by considering that when a particle squeezes part of a vesicle through the right opening of the cavity, it can only keep on transporting the vesicle to the next cell if its axis of propulsion forms an angle, $|\phi|$, with the x axis that is smaller than that formed by the normal to the slanted side of the cavity. This constraint does not apply when a particle is attempting to transport a vesicle through the left opening. In this case, a particle can complete the translocation process independently of its orientations as it faces a flat vertical wall, rather than a slanted wall on its way back. When that happens, the particle can still complete the translocation by dragging part of the  vesicle vertically as it slides along the flat wall for any value of $|\phi|>0$.

\section{Conclusions}
To conclude, we studied the interplay of elastic and active forces in an active vesicle. We measured the shape fluctuations and motility of a soft two-dimensional vesicle containing active Brownian particles in free space, in confined cavities and inside an asymmetric periodic channel. When unconstrained, we observe large positive curvature fluctuations which are suppressed upon increasing the density of active particles inside the vesicle. When confined within a spherical cavity containing a small opening, the average escape time of the vesicle has a non-monotonic dependence on the number of active particles, suggesting an optimal particle density for enhanced vesicle translocation. When placed  inside an asymmetric periodic channel the motion of the vesicle rectifies as long as its diameter is larger than the size of the cavity's smallest constriction. 

Our simulations clearly show how vesicle rectification emerges from  the complex coupling between the active and elastic forces of our model and is not a mere mere result of active particles pushing and sliding against the triangular asymmetric cavity~\cite{ghosh2013self}.  In fact, for low particle packing fraction, where it is easy to induce large curvature fluctuations on the vesicle, rectification is determined by the number of ways particles can push the vesicle across the asymmetric cavity once the translocation process is initiated. At moderate to large packing fractions, high curvature fluctuations, necessary for a translocation event to occur, can only develop as a result of the collective behavior of the particles which tend to more likely cluster near regions with the largest curvature.

It is quite remarkable that this model, which could be considered as what is possibly the simplest active model of a cell, can generate a sufficient level of complexity to produce behavior, such as ratchetaxis of a living cell when placed in asymmetric confining channels~\cite{le2020ratchetaxis}.

\section*{Acknowledgements}
A.C. acknowledges financial support from the National Science Foundation under Grant No. DMR-2003444. The authors acknowledge Albert Chen for a careful reading of the manuscript.

\bibliography{rsc} 
\bibliographystyle{apsrev4-1} 

\section{Appendix}
\subsection{Mean Squared Displacement of soft vesicle containing active particles}\label{msd_appendix}
\begin{figure*}[t]
\centering
\includegraphics[width=0.8\textwidth]{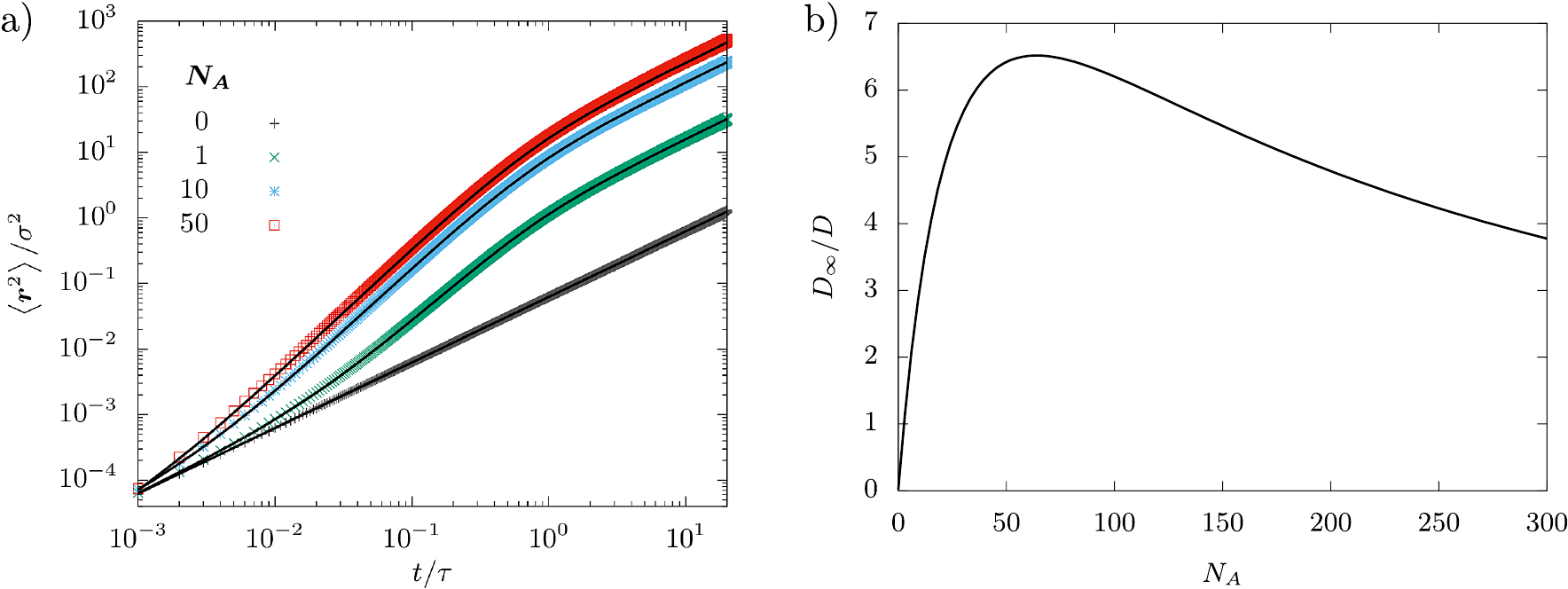}
 \caption{a) Mean squared displacement of center of mass of a vesicle containing active particles. With no active particles, the motion remains diffusive at all times. With a finite number of active particles enclosed, the vesicle has a ballistic regime at $t\approx\tau$ and is diffusive for $t\gg\tau$. The solid black lines are the closed form expression from Eq. \ref{correlate}. b) The long time diffusion constant (Eq. \ref{longd}) is non-monotonic and attains maximal value when the number of active particles $N_A$ in the vesicle equals the number of colloids forming the flexible boundary of the vesicle $N$. Here, $N=64$.}
 \label{msd}
\end{figure*} 

In this section, we evaluate the closed form expression for the mean squared displacement of the center of mass of the soft vesicle. This calculation follows the procedure adopted for run-and-tumble active particles within a vesicle~\cite{paoluzzi2016shape}.
The center of mass of the vesicle is defined as,
\begin{equation}
\bm{r}_{\rm cm}=\frac{1}{N+N_A}\left(\sum_{i=1}^N\bm{r}_i+\sum_{j=1}^{N_A}\bm{r}_j\right),
	\end{equation}
	where $\bm{r}_i,\bm{r}_j$ denote the positions of the particles in the flexible boundary and the enclosed active colloids respectively.
The velocity of the center of mass $\bm{v}=d\bm{r}/dt$ can be written explicitly in terms of the constituent particle velocities as,
\begin{equation}
\bm{v}_{\rm cm}=\frac{1}{N+N_A}\left(\sum_{i=1}^N\bm{v}_i+\sum_{j=1}^{N_A}\bm{v}_j\right).
\end{equation}

\noindent The dynamics of the particle velocities is encoded in the Brownian equations of motion in Eq. \ref{langevin} of the main text. The total force on the center of mass is the sum of all forces on all particles and given the symmetry of the problem the interaction forces between colloids add to zero~\cite{paoluzzi2016shape}.
We integrate the equations of motion of the colloidal particles to find the position of the center of mass,
\begin{equation}
\begin{split}
	\bm{r}_{\rm cm}(t)=&\frac{1}{N+N_A}\;\bigg(\sqrt{2D}\;\sum_{i=1}^N\int_0^t\bm{\xi}_i(t^\prime)\;dt^\prime\\
 +&\sqrt{2D}\;\sum_{j=1}^{N_A}\int_0^t\bm{\xi}_j(t^\prime)\;dt^\prime \\
 +&v_p\;\sum_{j=1}^{N_A}\int_0^t\bm{n}_j(t^\prime)\;dt^\prime\bigg).
 \end{split}
	\end{equation}

\noindent The mean squared displacement of the vesicle is,
\begin{equation}\label{correlate}
\begin{split}
	\left<\bm{r}_{\rm cm}^2(t)\right>=&\frac{1}{(N+N_A)^2}\times\\
 &\Bigg(2D\;\sum_{i=1}^N\int_0^t\int_0^t\left<\bm{\xi}_i(t^\prime)\cdot\bm{\xi}_i(t^{\prime\prime})\right>dt^\prime dt^{\prime\prime}\\
 &+2D\;\sum_{i=j}^{N_A}\int_0^t\int_0^t\left<\bm{\xi}_j(t^\prime)\cdot\bm{\xi}_j(t^{\prime\prime})\right>dt^\prime dt^{\prime\prime}\\
				&+v_p^2\;\sum_{j=1}^{N_A}\int_0^t\int_0^t\left<\bm{n}_j(t^\prime)\cdot\bm{n}_j(t^\prime)\right>\;dt^\prime dt^{\prime\prime}\Bigg).
	\end{split}
	\end{equation}
Given that noise on the colloidal particles is not correlated with each other and that the system is confined to two dimensions, we have $\left<\bm{\xi}_i(t^\prime)\cdot\bm{\xi}_i(t^{\prime\prime})\right>=2\;\delta(t^\prime-t^{\prime\prime})$. The first two sets of integrals in Eq. \ref{correlate} can be done employing the Dirac delta function. The third set of integrals involving the time-correlation of the orientation unit vector is evaluated following Hagen et al.~\cite{hagen2009non}. The ensemble of active particles considered in~\cite{hagen2009non} has a fixed initial direction of self-propulsion, which contributes direction specific terms to the mean squared displacement. However, since we numerically calculate the mean squared displacement of the vesicle along a trajectory, these terms average out to zero. 
Following the evaluation of the integrals in Eq. \ref{correlate}, we are able to write the mean squared displacement of the center of mass of the vesicle confined to two dimensions as,
\begin{equation}\label{msd_full}
\left<\bm{r}_{\rm cm}^2(t)\right>=\frac{4D\;t}{N+N_A}+\frac{2\;Pe^2\sigma^2N_A}{9\;(N+N_A)^2}\left(e^{-D_r\;t}+D_r\;t-1\right).
\end{equation}

\noindent The numerical results are tested against this expression in Fig.~\ref{msd}(a). The long time diffusion constant $D_\infty$ defined as, 
\begin{equation}\label{longd}
D_\infty=\lim_{t \to \infty} \frac{1}{4t}\left<\bm{r}_{\rm cm}^2(t)\right>=\frac{D}{N+N_A}+\frac{Pe^2\sigma^2N_AD_r}{18(N+N_A)^2},
\end{equation}
is non-monotonic as a function of $N_A$  (See Fig. \ref{msd}b). The number of active particles $N_A^0$ needed to attain the maximal $D_\infty$ is calculated by setting the derivative of $D_\infty$ with respect to $N_A$ to be zero. This condition can be simplified to,
\begin{equation}
\frac{N_A^0}{N+N_A^0}=\frac{1}{2}-\frac{3}{Pe^2}.
\end{equation}
For $Pe\rightarrow\infty$, $N_A^0=N$.  

\subsection{Preferential motility of a vesicle stuck between sloped walls}\label{angle_appendix}
\begin{figure}[h]
	\centering
	\includegraphics[width=0.5\textwidth]{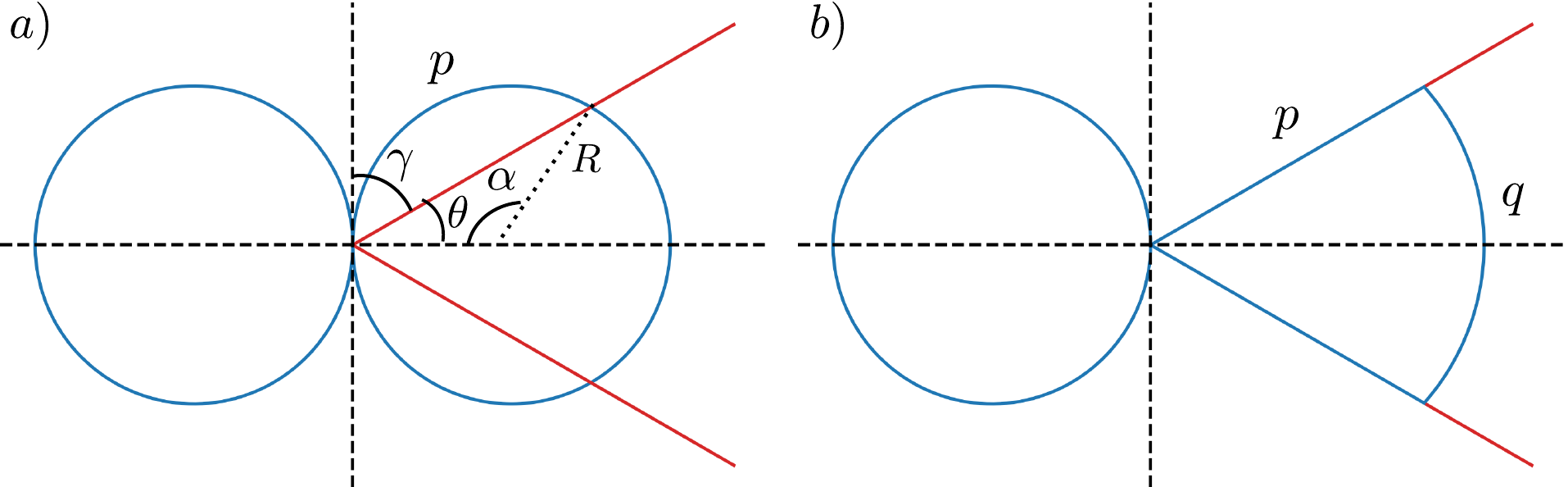}
	\caption{Geometries used to calculate $P^L(\gamma)$. We ignore the finite pore width. In (b), we assume the length of the ring in contact with the wall $p$ to be the same as the arc length in (a).}\label{slant walls}
	\end{figure}

In this section, we estimate the probability of leftward transitions $P^L(\gamma)$ (Eq. 3 of main text) for the system in the inset of Fig.~3 of main text. We approximate the soft vesicle as  two rings of equal radius $R$ whose perimeters are $2\pi R$ each. We also ignore the finite pore width and set it to zero to allow for a simple calculation. See Fig.~\ref{slant walls}(a). The angle $\gamma$ quantifies the slope of the slanted wall with respect to the $y-$ axis. The wall bisects the right side ring at two points. The angle bisected by the arc between these two points is $\alpha$, and the arc length is $p=R\alpha$. We assume that for $N_A\gg1$, the ring conforms to the wall as in Fig.~\ref{slant walls}(b), where the portion of the ring in contact with the wall is taken to be $p$. The length of curved portion of the right ring is $q=2\pi R-2p=2 R(\pi-\alpha)$. From the geometry of the system, we have $\alpha=\pi-2\theta$ and $\theta=\pi/2-\gamma$. In the high $Pe$ limit, all particles are expected to be at the vesicle boundary. The number of active particles at the vesicle boundary is proportional to the length of the vesicle, and the force on the ring scales with the number of active particles. The probability of leftward transition is then,
$$P^L=\frac{2\pi R}{2\pi R+q}.$$
While all the active particles on the left ring push the vesicle leftwards, only the active particles on the free curved portion of the right ring push the vesicle rightwards. Using the geometric relations discussed above, $P^L$ as a function of the angle $\gamma$ can be found to be Eq. \ref{pressure diff} of the main text. 

\end{document}